\documentclass[aps,prb,reprint,superscriptaddress,showpacs, citeautoscript]{revtex4-1}

\usepackage{graphicx}             
\usepackage{amsmath}              
\usepackage{amssymb}              
\usepackage{braket}              
\usepackage[load=prefixed]{siunitx}
\usepackage[usenames,dvipsnames,svgnames,table]{xcolor}
\usepackage[english]{babel}
\usepackage{ifthen}
\usepackage[hidelinks]{hyperref}
\usepackage{ulem}
\usepackage{blindtext}
\usepackage[caption=false]{subfig}    
\usepackage[miktex]{gnuplottex}   
\usepackage{epstopdf}	          
\usepackage{multirow}
\usepackage[capitalize]{cleveref}
\usepackage{todonotes}			

\definecolor{dark-red}{HTML}{8B0000}
\definecolor{dark-green}{HTML}{006400}

\newcommand{\mpoint}{\,.}					
\newcommand{\mkomma}{\, ,}	
\newcommand{\figref}[1]{\mbox{Fig. \ref{#1}}}	
\newcommand{\mexp}[1]{\mathrm{e}^{#1}}		
\newcommand{\person}[1]{\textit{#1}}        

\newcommand{\secref}[1]{\mbox{\cref{#1}}}	  
\newcommand{\tabref}[1]{\mbox{Tab. \ref{#1}}}     

\newcommand{\sNi}{SrCu$_{1-x}$Ni$_x$O$_2$\xspace}
\newcommand{\sZn}{SrCu$_{1-x}$Zn$_x$O$_2$\xspace}

\begin{document}
\title{The effect of different in-chain impurities on the magnetic properties of the spin chain compound SrCuO$_2$ probed by NMR}
\date{\today}
\author{Yannic Utz}
\affiliation{IFW Dresden, Helmholtzstra{\ss}e 20, 01069 Dresden, Germany}
\author{Franziska Hammerath}
\affiliation{IFW Dresden, Helmholtzstra{\ss}e 20, 01069 Dresden, Germany}
\affiliation{Institute for Solid State Physics, Dresden Technical University, TU-Dresden, 01062 Dresden, Germany}
\author{Roberto Kraus}
\affiliation{IFW Dresden, Helmholtzstra{\ss}e 20, 01069 Dresden, Germany}
\author{Tobias Ritschel}
\affiliation{Institute for Structure Physics, Dresden Technical University, TU-Dresden, 01062 Dresden, Germany}
\author{Jochen Geck}
\affiliation{Institute for Structure Physics, Dresden Technical University, TU-Dresden, 01062 Dresden, Germany}
\author{Liviu Hozoi}
\affiliation{IFW Dresden, Helmholtzstra{\ss}e 20, 01069 Dresden, Germany}
\author{Jeroen van den Brink}
\affiliation{IFW Dresden, Helmholtzstra{\ss}e 20, 01069 Dresden, Germany}
\affiliation{Institute of Theoretical Physics, Dresden Technical University, TU-Dresden, 01062 Dresden, Germany}
\author{Ashwin Mohan}
\altaffiliation[Present address: ]{Institute of Chemical Technology, Mumbai-400019, India.}
\affiliation{IFW Dresden, Helmholtzstra{\ss}e 20, 01069 Dresden, Germany}
\author{Christian Hess}
\affiliation{IFW Dresden, Helmholtzstra{\ss}e 20, 01069 Dresden, Germany}
\author{Koushik Karmakar}
\altaffiliation[Present address: ]{IFW Dresden, Helmholtzstra{\ss}e 20, 01069 Dresden, Germany}
\affiliation{Indian Institute of Science Education and Research, Pune, Maharashtra-411008, India}
\author{Surjeet Singh}
\affiliation{Indian Institute of Science Education and Research, Pune, Maharashtra-411008, India}
\author{Dalila Bounoua}
\affiliation{SP2M-ICMMO, UMR-CNRS 8182, Université Paris-Sud, Université Paris-Saclay, 91405 Orsay, France}
\author{Romuald Saint-Martin}
\affiliation{SP2M-ICMMO, UMR-CNRS 8182, Université Paris-Sud, Université Paris-Saclay, 91405 Orsay, France}
\author{Loreynne Pinsard-Gaudart}
\affiliation{SP2M-ICMMO, UMR-CNRS 8182, Université Paris-Sud, Université Paris-Saclay, 91405 Orsay, France}
\author{Alexandre Revcolevschi}
\affiliation{SP2M-ICMMO, UMR-CNRS 8182, Université Paris-Sud, Université Paris-Saclay, 91405 Orsay, France}
\author{Bernd B\"uchner}
\affiliation{IFW Dresden, Helmholtzstra{\ss}e 20, 01069 Dresden, Germany}
\affiliation{Institute for Solid State Physics, Dresden Technical University, TU-Dresden, 01062 Dresden, Germany}
\author{Hans-Joachim Grafe}
\affiliation{IFW Dresden, Helmholtzstra{\ss}e 20, 01069 Dresden, Germany}

\begin{abstract}
The $S=1/2$ Heisenberg spin chain compound SrCuO$_2$ doped with different amounts of nickel (Ni), palladium (Pd), zinc (Zn) and cobalt (Co) has been studied by means of Cu nuclear magnetic resonance (NMR). Replacing only a few of the S=1/2 Cu ions with Ni, Pd, Zn or Co has a major impact on the magnetic properties of the spin chain system. In the case of Ni, Pd and Zn an unusual line broadening in the low temperature NMR spectra reveals the existence of an impurity-induced local alternating magnetization (LAM), while exponentially decaying spin-lattice relaxation rates $T^{-1}_1$ towards low temperatures indicate the opening of spin gaps. A distribution of gap magnitudes is proven by a stretched spin-lattice relaxation and a variation of $T^{-1}_1$ within the broad resonance lines. These observations depend strongly on the impurity concentration and therefore can be understood using the model of finite segments of the spin $1/2$ antiferromagnetic Heisenberg chain, i.e. pure chain segmentation due to $S=0$ impurities. This is surprising for Ni as it was previously assumed to be a magnetic impurity with $S=1$ which is screened by the neighboring copper spins. In order to confirm the $S=0$ state of the Ni, we performed x-ray absorption spectroscopy (XAS) and compared the measurements to simulated XAS spectra based on multiplet ligand-field theory. Furthermore, Zn doping leads to much smaller effects on both the NMR spectra and  the spin-lattice relaxation rates, indicating that Zn avoids occupying Cu sites. For magnetic Co impurities, $T^{-1}_1$ does not obey the gap like decrease, and the low-temperature spectra get very broad. This could be related to the increase of the N\'{e}el temperature which was observed by recent $\mu$SR and susceptibility measurements \cite{Karmakar2017}, and is most likely an effect of the impurity spin $S\neq0$.
\end{abstract}


\maketitle

\section{Introduction}
Much attention has been paid to the issue of impurities in low-dimensional spin systems. Their strong impact on the ground state as well on the excitations leads to effects which are interesting in their own right and which are also of some interest for the understanding of high temperature superconductivity. 
Materials which can be described by the model of the $S=1/2$ antiferromagnetic Heisenberg chain are thereby particularly interesting, as this model is exactly solvable and is, therefore, often used as an archetype for low-dimensional quantum magnets in theory.
Despite its simplicity, it shows a very unusual behavior. 
The ground state of this model is an example of a highly entangled many-body quantum state, which is characterized by a lack of long-range order even at absolute zero. Its elementary excitations are exotic quasiparticle excitations with fractional quantum numbers, the $S=1/2$ spinons, which can be excited with infinitely low energy, i. e. the excitation spectrum has no energy gap to the ground state \cite{Auerbach1994,Schollwoeck2004,Lieb1961,Cloizeaux1962}. 
Small perturbations induced by impurities or by interchain interactions are expected to lead to gaps in the excitation spectra or to three-dimensional (3D) long-range ordering \cite{Affleck1989,Eggert1992,Affleck1994,Kojima1997,Laukamp1998, Eggert2002}.

The closely related cuprate compounds  SrCuO$_2$ and Sr$_2$CuO$_3$ are known to be among the best realizations of the 1D \mbox{$S=1/2$} antiferromagnetic Heisenberg model. There, the chains are realized by corner sharing CuO$_4$ plaquettes with \mbox{$S=1/2$} on the copper site, which are mainly interacting along one crystallographic axis with a large exchange coupling of about $J \sim \SI{2000}{K}$ \cite{Motoyama1996, Suzuura1996, Eisaki1997, Zaliznyak1999, Zaliznyak2004}.
While Sr$_2$CuO$_3$ implements a single chain structure, SrCuO$_2$ contains double chains coupled by a weak and fully frustrated interchain coupling $|J^\prime|  \sim \num{0.1} J$ \cite{Rice1993,Motoyama1996}.
Weak static magnetism occurs only below  $T_N=\SI{2}{K}$ and $T_N = \SI{5.4}{K}$ in SrCuO$_2$ and Sr$_2$CuO$_3$, respectively \cite{Keren1993, Kojima1997, Matsuda1997}, which is low compared to the much larger exchange coupling $J$.

However, recent studies on doped variants of these compounds revealed the vulnerability of the originally gapless spinon excitation spectrum \cite{Zaliznyak2004, Takigawa1996} against the influence of impurities and disorder.
$^{63}$Cu nuclear magnetic resonance and transport studies showed that doping Ca on the Sr site outside the chains breaks the integrability of the model and opens a spin gap of similar size in both compounds, which has been attributed to structural distortions and a concomitant bond disorder \cite{Hammerath2011a, Hlubek2011, Hammerath2014, Mohan2014}.
Inelastic neutron scattering disclosed a striking impact of minor concentrations of in-chain nickel impurities on the low-energy spin dynamics of the double chain compound SrCuO$_2$ \cite{Simutis2013}. 
The authors report the emergence of a spin pseudogap of the order of $\Delta\approx\SI{90}{K}$ by replacing only $\SI{1}{\percent}$ of the \mbox{$S=1/2$} copper ions with nickel impurities. 
According to their interpretation, nickel carries a \mbox{$S=1$} which is fully screened. Therefore, the nickel ions effectively act as $S=0$ impurities and basically cut the chains into segments of varying finite length $l$, 
which show finite-size spin gaps with magnitudes proportional to $1/l$ \cite{Eggert1992}. On average, this leads to the observed pseudogap behavior on a macroscopic scale.
With NMR spin-lattice relaxation measurements, it was possible to evidence this distribution of finite-size spin gaps also in Ni-doped Sr$_2$CuO$_3$, not only by a distribution of decreasing spin-lattice relaxation rates at low temperatures but also by the doping dependence of the onset temperature of this decrease \cite{Utz2015}. However, the NMR spectra showed a suppression of the impurity-induced staggered paramagnetic response with increasing doping level which was argued to be due to the increasing gap magnitude \cite{Utz2015}.

In this paper, we show NMR results on SrCuO$_2$ doped with nickel (Ni), palladium (Pd), zinc (Zn), and cobalt (Co). All of these impurities are assumed to replace copper ions and, therefore, to produce chain defects. Pd and Zn are $S=0$ impurities \cite{Sirker2007,Kojima2004,Kawamata2008,Alloul2009} and are, therefore, expected to cut the chains into segments of finite lengths and to lead to a similar behavior as Ni doping. 
Ni is special in so far as it is not yet sure if it carries $S=0$ or $S=1$ in the chain. Previously, it has mostly assumed to be an $S=1$ impurity as in the 2D cuprates \cite{Alloul2009}.
In this case, the spin should be screened by the surrounding Cu spins and, therefore, acts effectively as a spin 0 impurity \cite{Eggert1992}.
However due to the square planar arrangement of the oxygen ions in the chain cuprates, the Ni impurity could also be in a low spin state as a native $S=0$ impurity \cite{Nishida1979, Chattopadhyay2011, Matsuda1999} similar to the compound BaNiO$_2$ which is structurally close to SrCuO$_2$ \cite{Chattopadhyay2011, Matsuda1999, Krischner1971}.
Around spin 0 impurities, a local alternating magnetization (LAM) arises (see below for details), which leads to a magnetic broadening of the resonance lines.
For a screened $S=1$ impurity, the LAM is not only determined by a backscattering contribution but also by the Kondo screening, which changes the shape of the NMR spectra \cite{Rommer2000}.
Thus by comparing the NMR spectra of a Ni-doped sample with the ones of a sample doped with $S=0$ impurities such as Pd, we clarified this issue. Further XAS measurements confirmed that Ni is in a low spin $S=0$ configuration.
Zinc, which is also a scalar impurity, could in principle be used for this purpose, too. However, it turned out that the actual doping level of the Zn-doped samples is much smaller than the nominal one and that Zn avoids occupying copper sites other than in the layered cuprate compounds.
The spin state of Co in the chain is also not clear up to now. As Ni, it could be in a high or low spin state. However, in the case of Co this means either $S=3/2$ or $S=1/2$. Both should lead to a differing behavior as compared to doping with $S=0$ impurities \cite{Eggert1992}.

The paper is structured as following: After an introduction to the samples and to the experimental methods, measurements on the different dopings are shown and discussed one after another. At first, a detailed discussion of the results on the Ni-doped compounds, which have been most intensively studied,  is given. Afterwards, the measurements on a Pd-doped sample of SrCuO$_2$ are shown and compared to the Ni-doped case, which primarily allows conclusions about the latter. Then, the Zn-doped compounds are considered, where the comparison to Ni doping is again of importance to allow for conclusions. At last, the case of Co doping is discussed in view of its differences to the other impurities.

\section{Experimental Detail}
\subsection{Samples}
The measurements were performed on high purity single
crystals of SrCuO$_2$, SrCu$_{1-x}$Ni$_x$O$_2$ (x = 0.0025, 0.005 and 0.01), SrCu$_{1-x}$Zn$_x$O$_2$ (x = 0.01 and 0.02), SrCu$_{0.99}$Pd$_{0.01}$O$_2$ and SrCu$_{0.99}$Co$_{0.01}$O$_2$ (see \tabref{tab:samples} for an overview and for the short names used hereafter). The samples were prepared using the traveling solvent floating zone (TSFZ) method using starting materials of at least \SI{99.99}{\percent} purity \cite{Revcolevschi1999,Moh2014,Saint-Martin2015,Karmakar2014}. The
high quality of the crystals has been checked by x-ray diffraction (phase determination) and energy-dispersive x-ray spectroscopy (chemical composition) measurements. After determining the orientation with a Laue camera, preferably cubic shaped pieces were cut out of the crystals to have samples with the edges aligned along the crystallographic axes. Typical sample sizes were \SI{2}{mm} to \SI{4}{mm} in length and \SI{0.5}{mm} to \SI{1}{mm} along the other two dimensions.

\begin{table}
	\resizebox{0.8\columnwidth}{!}{%
		\begin{tabular}{|c|c|c|c|c|}
			\hline \rule[-2ex]{0pt}{5.5ex} Short Name & Material (nominal) &  Grown By\\ 
			\hline \rule[-2ex]{0pt}{5.5ex}  N0 & SrCuO$_2$  & R. Saint-Martin\textsuperscript{1} \\ 
			\hline \rule[-2ex]{0pt}{5.5ex}  N0.25 & SrCu$_{0.9975}$Ni$_{0.0025}$O$_2$  & R. Saint-Martin\textsuperscript{1} \\ 
			\hline \rule[-2ex]{0pt}{5.5ex}  N0.5 & SrCu$_{0.995}$Ni$_{0.005}$O$_2$ &  R. Saint-Martin\textsuperscript{1} \\ 
			\hline \rule[-2ex]{0pt}{5.5ex}  N1 & SrCu$_{0.99}$Ni$_{0.01}$O$_2$  &  A. Mohan\textsuperscript{2} \\ 
			\hline \rule[-2ex]{0pt}{5.5ex}  Z1 & SrCu$_{0.99}$Zn$_{0.01}$O$_2$ &  K. Karmakar\textsuperscript{3} \\ 
			\hline \rule[-2ex]{0pt}{5.5ex}  Z2 & SrCu$_{0.98}$Zn$_{0.02}$O$_2$ &  K. Karmakar\textsuperscript{3} \\
			\hline \rule[-2ex]{0pt}{5.5ex}  P1 & SrCu$_{0.99}$Pd$_{0.01}$O$_2$ &  D. Bounoua\textsuperscript{1} \\ 
			\hline \rule[-2ex]{0pt}{5.5ex}  C1 & SrCu$_{0.99}$Co$_{0.01}$O$_2$ & K. Karmakar\textsuperscript{3} \\ 
			\hline 
		\end{tabular}}
		\caption{Overview of used samples and their producers. The samples were grown by different persons from different groups. The superscripts refer to the following affiliations:  \textsuperscript{1}SP2M-ICMMO UMR-CNRS, Orsay, France -- \textsuperscript{2}IFW Dresden, Germany -- \textsuperscript{3}IISER, Pune, India }
		\label{tab:samples}
	\end{table}
		 
\subsection{NMR Measurements}
Three types of NMR measurements are presented and discussed in this paper.
\textbf{Field-swept NMR spectra} have been obtained and \textbf{Cu NMR spin-lattice relaxation measurements} have been performed for each sample at temperatures between \SI{4.2}{K} and \SI{300}{K}.
The spin-lattice relaxation measurements have been performed on the center of the $^{63}$Cu high-field satellite as the mainline is affected by an additional peak (see below).
On the samples N1 and Z2, additional \textbf{frequency-dependent spin-lattice relaxation measurements} have been performed at different positions within the broad $^{63}$Cu high-field satellite.
The field-swept NMR spectra have been measured with a fixed frequency of \SI{80}{MHz} by varying the magnetic field $H$ along the crystallographic $b$ axis in steps of \SI{1}{mT} and plotting the integrated echo intensity of a \ang{90}-\ang{180}-pulse-sequence.
The spin-lattice relaxation measurements have been performed using the inversion recovery method in magnetic fields close to \SI{7}{T}. The recovery curves of the nuclear magnetization have been fit to the standard recovery function for magnetic relaxation of a satellite transition of $I=3/2$ nuclei \cite{McDowell1995} modified by a stretching exponent $\lambda \leq 1$ to account for a distribution of spin-lattice relaxation rates \cite{Johnston2006} at low temperatures: 
\begin{align}
M_z(t) = & M_0\Big[1-f\Big(0.4\mexp{-(6t/T_1)^\lambda}+0.5\mexp{-(3t/T_1)^\lambda} \nonumber\\
         & +0.1\mexp{-(t/T_1)^\lambda}\Big)\Big]\mpoint
\end{align}
$M_0$ is the equilibrium value of the nuclear magnetization and $f$ is ideally $2$ for a complete inversion.
The spectra shown for guidance together with the frequency-dependent spin-lattice relaxation measurements in \figref{fig:freqdepT1} were obtained at a fixed
field $H$ by sweeping the frequency and adding the Fourier transforms of the echo signals (frequency step and sum method \cite{Clark1995}).

\section{Results and Discussion}
\subsection{Nickel-Doped Samples}
\label{sec:NiDoping}

\subsubsection{NMR Spectra}
\begin{figure}  
	\includegraphics[width=\columnwidth]{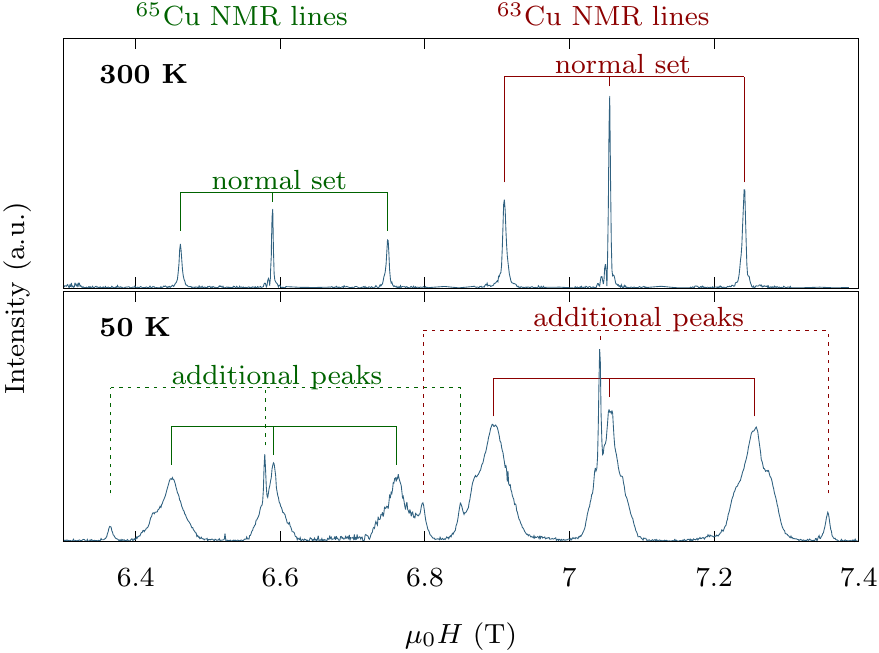}
	\caption{Complete Cu NMR spectra of SrCu$_{0.99}$Ni$_{0.01}$O$_2$ at \SI{300}{K} and \SI{50}{K} as examples. They have been obtained with a fixed frequency of \SI{80}{MHz} by varying the magnetic field $H$ while keeping it parallel to the crystallographic $b$ axis (perpendicular to the chains). The normal set of Cu NMR lines and the additional set due to a local lattice distortion around the nickel impurities are marked.}
	\label{fig:spectrum1Ni50K}
\end{figure}

\begin{figure}  
	\centering
	\includegraphics[width=\columnwidth]{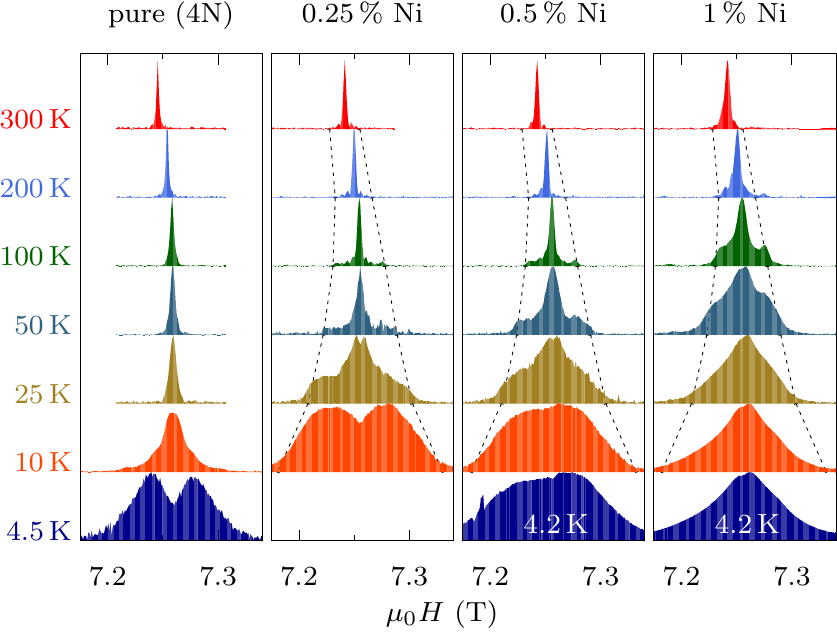}
	\caption{$^{63}$Cu high-field satellite of SrCuO$_2$ and \sNi at different temperatures measured with a fixed frequency of $\SI{80}{MHz}$ by varying the magnetic field $H||b$ (perpendicular to the chains). The black dashed lines (more precise: their intersections with the base lines of the spectra) indicate the $1/\sqrt(T)$-behavior, which is expected from the model of semi-infinite chains with \mbox{$\sqrt(T)\Delta H/2H_0 = 0.0326$}.}
	\label{fig:highfieldsat_comp_SrCuO2_Ni}
\end{figure}

\figref{fig:spectrum1Ni50K} shows the complete Cu NMR spectra of N1 at \SI{300}{K} and \SI{50}{K} as examples. At \SI{300}{K}, six resonance lines are visible --- one central line and two quadrupolar split satellites for $^{63}$Cu and $^{65}$Cu, respectively. Essentially, this spectrum does not differ from the high temperature spectrum of the pure compound. However, the satellites slightly broaden with increasing doping level (see also \figref{fig:highfieldsat_comp_SrCuO2_Ni}), in contrast to the mainline which keeps its width. This is a consequence  of the increasing structural disorder induced upon doping, which leads to slight variations of the electric field gradient (EFG) and, therefore, of the quadrupolar splitting.
Towards low temperatures, these resonance lines broaden and obtain some structure. In the \SI{50}{K} spectrum shown in \figref{fig:spectrum1Ni50K}, one can see that  mainlines and satellite lines are equally affected, which proves that the broadening is of magnetic origin.

Before studying this magnetic broadening in more detail for different doping levels and temperatures, another feature should be mentioned.
At low temperatures, an additional set of resonance lines can be observed for the Ni-doped samples. They are marked by dashed lines in the \SI{50}{K} spectrum shown in \figref{fig:spectrum1Ni50K}. The ratio of  the integrated intensity of the $^{63}$Cu high-field satellite of these additional peaks to the integrated intensity of the "normal" $^{63}$Cu high-field satellite  is independent of temperature, but depends on the Ni content\footnote{The additional peaks are not observed at high temperatures, because the "normal" lines are very narrow there, which means that their peak intensity is high. Therefore, the spectra were obtained with lower signal-to-noise ratio as compared to the spectra at lower temperatures.}. The ratio is \SI{2.2\pm0.7}{\percent}, \SI{3.4\pm1.0}{\percent}, and \SI{5.6\pm1.0}{\percent} for \SI{0.25}{\percent}, \SI{0.5}{\percent}, and \SI{1}{\percent} of Ni doping, respectively. Thus, it is monotonically growing with the doping level. This and the observed larger quadrupolar splitting of the additional satellites
leads to the conclusion that these additional peaks are due to Cu sites close to Ni impurities. The Ni defects may cause local changes in the EFG and, therefore, a larger quadrupolar splitting.
An open question is, however, why this happens only with nickel doping and not for other impurities. Furthermore, the emergence of this distortions is quite surprising as SrNiO$_2$ crystallizes with the same lattice structure and alsmost the same lattice parameters as SrCuO$_2$ \cite{Pausch1976}.
The additional peaks overlap with the "normal" resonance lines. Therefore in the following, only the $^{63}$Cu high-field satellite which is not disturbed by additional lines will be studied in detail.

\figref{fig:highfieldsat_comp_SrCuO2_Ni} shows the $^{63}$Cu high-field satellite for the Ni-doped samples and for the pure compound as a reference. The line of the pure compound stays narrow over a wide temperature range. Only at $T\leq \SI{10}{K}$, the line broadens and splits. This is most probably due to increasing spin-spin correlations related to the close-by phase transition to 3D order at $T_N \approx 2K$ \cite{Zaliznyak1999, Matsuda1995}.
As already mentioned, the spectra of the Ni-doped samples show a magnetic broadening towards low temperatures. They develop shoulder structures and a splitting of the tip very similar to the NMR spectra of the closely related spin chain compound Sr$_2$CuO$_3$ doped with nickel \cite{Utz2015}. This is typical for a local alternating magnetization (LAM) around impurities.

\begin{figure}
	\centering
	\includegraphics[width=\columnwidth]{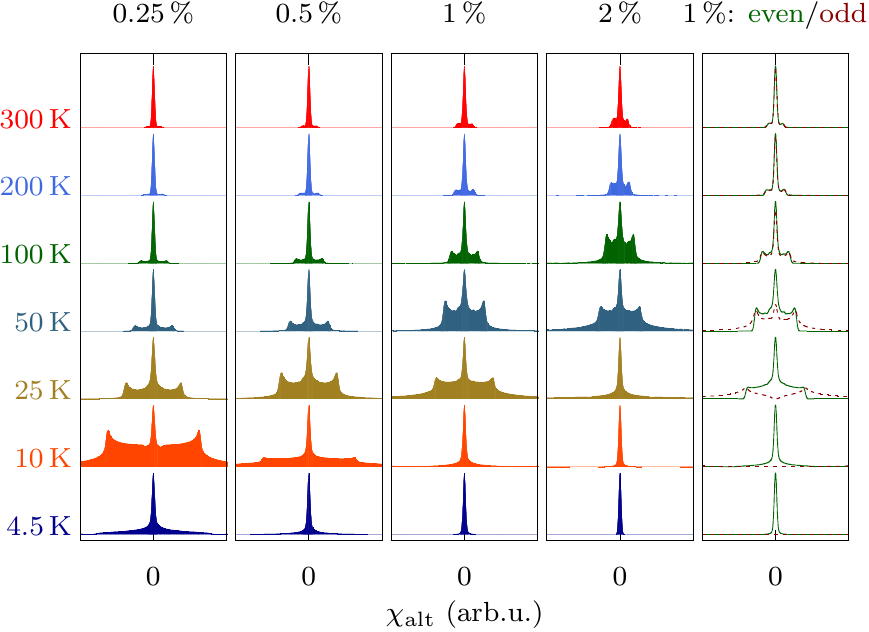}
	\caption{Spectra simulated based on the model of finite-sized chain segments (see text). The spectra are each normalized to to their maximum. The last panel shows the contribution of chain segment with even and odd number of sites separately.}
	\label{fig:simulated_LAM_finite}
\end{figure}

A LAM has been observed previously in undoped Sr$_2$CuO$_3$, where it has been explained by open chain ends due to excess oxygen \cite{Takigawa1997a,Boucher2000,Sirker2009}. These chain ends break the translational invariance of the spin chain  and lead to a local alternating susceptibility  [$\chi_\mathrm{alt}(x)$], which gives rise to a LAM in a magnetic field.
It could be modeled based on the assumption of semi-infinite chains \cite{Eggert1995}, which predicts a LAM with a maximum at a certain distance $l=\num{0.48}J/T$  from the impurity \cite{Takigawa1997a} and an exponential decay for larger distances. Upon lowering the temperature, the maximum should shift further into the chain and should increase with $\chi_\mathrm{alt,max} \propto 1/\sqrt{T}$.
In the NMR spectra, this should cause a broad background with sharp edges, which broadens with decreasing temperature corresponding to $\Delta H \propto 1/\sqrt{T}$ independent of the amount of chain breaks \cite{Eggert1995, Takigawa1997a}. The intensity of the background should increase with decreasing temperature. 

We can identify the shoulder features as this broad background. As expected, their width is independent of the doping level, but their intensity increases with increasing impurity concentration and with decreasing temperature. 
The dashed lines in \figref{fig:highfieldsat_comp_SrCuO2_Ni} indicate the expected $1/\sqrt{T}$ behavior.\footnote{The $1/\sqrt(T)$ behavior was fitted to the width $\Delta H$ of the shoulder feature at $\SI{100}{K}$ and corresponds to $\sqrt(T)(\Delta H / 2H_0) = 0.0326$, which is a bit smaller than for Sr$_2$CuO$_3$ \cite{Takigawa1997a, Utz2015} due to differing hyperfine couplings. } The line shape clearly follows this trend down to \SI{25}{K}, \SI{50}{K}, and \SI{100}{K} for \SI{0.25}{\percent}, \SI{0.5}{\percent}, and \SI{1}{\percent} of Ni doping, respectively. At lower temperatures, however, the shoulder feature is smeared out. In the measurements at lowest temperatures, one can see that the resonance line is even getting narrower for higher doping levels --- an unusual behavior which has already been observed in Ni-doped Sr$_2$CuO$_3$ \cite{Utz2015}.

However, the smearing of the shoulder features and the suppression of the low-temperature linewidth with increasing doping level can be understood taking the finite size of the chain segments into account. At low temperatures, the LAM extends over the whole chain segment and, therefore, the number of sites gets important. As the ground state of segments with even number of sites is a singlet, its LAM comes from an excited state above the gap. Therefore, the amplitude of the LAM of even chain segments disappears exponentially as the temperature decreases \cite{Nishino2000, Sirker2009}. The doublet ground state of segments with an odd number of sites, however, leads to an increase of the amplitude of the LAM with decreasing temperature.
\person{Sirker} and \person{Laflorencie} obtained a formula for the alternating part of the local spin susceptibility of finite chain segments based on field theory \cite{Sirker2009}:
\begin{align}
\chi_l^\mathrm{alt} = &- \frac{c}{T} \left(\frac{\pi}{n+1}\right)^{1/2} \frac{\eta^{3/2}\left(\mexp{-\frac{\pi^2J}{2k_BTL}}\right)}{\theta_1^{1/2}\left(\frac{\pi l}{N+1},\mexp{-\frac{\pi^2 J}{4k_BTL}}\right)} \nonumber \\
& \cdot \frac{\sum_m m \sin[2\pi m l/(N+1)]\mexp{-\pi^2Jm^2/(2Lk_BT)}}{\sum_m \mexp{-\pi^2Jm^2/(2Lk_BT)}} \quad \mkomma 
\label{eq:LAM_finite_chain}
\end{align}
where $l=na$ is the position within the chain of length $L=Na$ with the lattice constant $a$, $c$ is a prefactor which was just set to $1$ here, $\eta(x)$ is the Dedekind eta function and $\theta_1(u,q)$ the elliptic theta function of the first kind. 
Based on this formula, we simulated spectra for distributions of chain segments with the probability\cite{Simutis2013} $P_L = x^2(1-x)^L$ to find a non-interrupted chain segment of length $L$  for given defect concentrations of $x = 0.0025, 0.005, 0.1,$ and $0.2$. The results are shown in \figref{fig:simulated_LAM_finite}.
The simulated  spectra resemble very much the measured ones in \figref{fig:highfieldsat_comp_SrCuO2_Ni}. At high temperatures (e.g. \SI{100}{K}), there are shoulder features which show an increasing intensity with increasing doping level. Towards lower temperatures, these shoulder features are degraded. The onset of this process depends, as in the experiment, on the doping level as the shape of the LAM depends now on the length of the individual chain segments. But in contrast to the experiment, the low temperature spectra are very narrow, as the even-length segments are frozen in their singlet ground states.
This can also be seen in the last panel, where the contribution of even and odd chain segments is displayed separately for a doping level of \SI{1}{\percent}. It shows furthermore, that the contribution of odd segments should become unobservable at lowest temperatures according to the model as their local susceptibility gets very large, which leads to a smearing of their spectral intensity over a very broad range such that it cannot be resolved anymore.
Therefore, the reasoning in Ref. \onlinecite{Utz2015} that the low-temperature linewidth is suppressed by the finite-size gap is valid only for the even chain segments. The odd ones just become unobservable.

But which additional aspect might be responsible for the broadening of the experimental low-temperature spectra as compared to the simulated ones based on a finite-size single chain model?
One reason could be the very small interchain coupling which also accounts for the 3D ordering of pure SrCuO$_2$ at $T_N\approx\SI{2}{K}$ \cite{Zaliznyak1999, Matsuda1995}.
The finite spin chain model does not only show a staggered response to a uniform field but also to a staggered field \cite{Eggert2002}. Therefore, a staggered magnetization in one chain can induce a staggered response in a neighboring one via the interchain interaction. This mechanism is closely connected to the formation of N\'{e}el order \cite{Eggert2002}. As both the broadening of the low-temperature spectra and the N\'{e}el order are promoted by the same mechanism within this scenario, the observation that both the width of the broad low-temperature lines and the N\'{e}el temperature\cite{Karmakar2015b} are suppressed with increasing doping level gives evidence for the validity of this scenario.

Another reason could be the nickel spin. Up to here, it cannot be excluded that the deviation from the finite-size single chain model is due to a possible nickel spin 1. The smearing of the shoulder feature might be just related to the screening of the impurity spin \cite{Rommer2000}. However within this scenario, it would be difficult to explain why the low-temperature linewidth should decrease with increasing doping level. There is no reason why the screening cloud should be suppressed when more impurities are screened. To clarify this issue, the results are compared to the measurements on a Pd-doped sample in \ref{sec:PdDoping}, as Pd is a $S=0$ impurity for sure. Furthermore, we conducted x-ray absorption spectroscopy (XAS) measurements, which follow in the next section \ref{sec:xas}.

\subsubsection{X-ray absorption spectroscopy}
\label{sec:xas}

\begin{figure}
	\centering
	\includegraphics[width=\columnwidth]{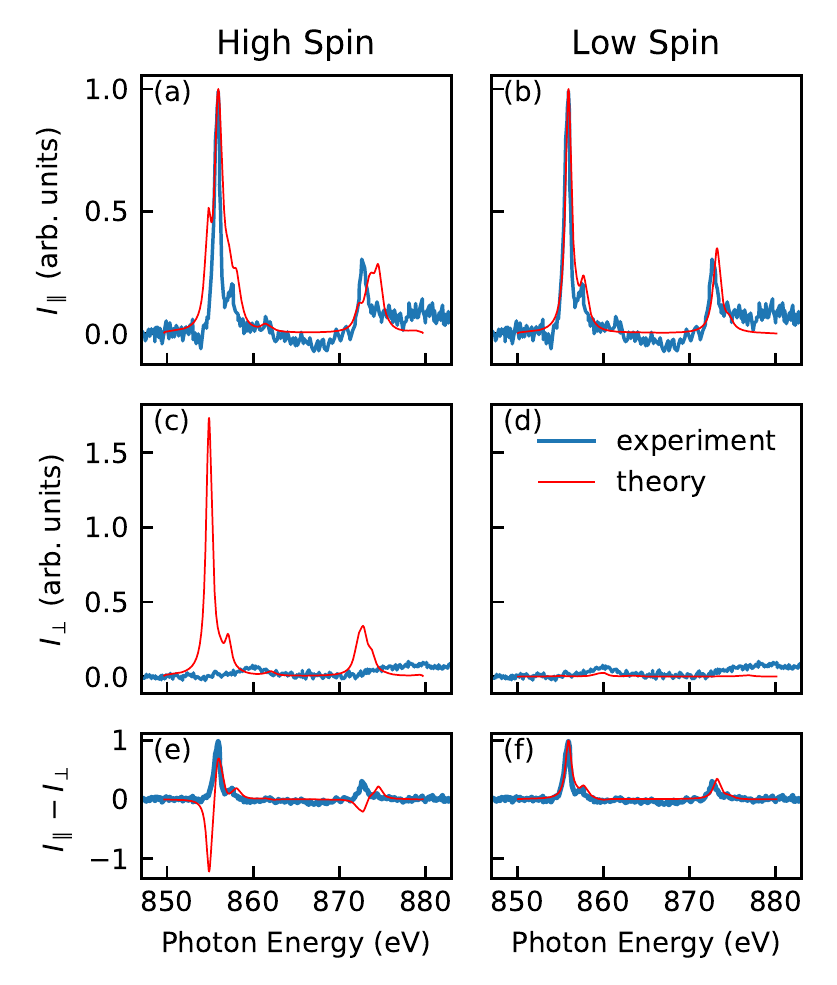}
	\caption{XAS at the Ni L$_{2/3}$-edge for a sample with 1\% Ni doping compared to multiplet calculations for a high spin (left panel: (a), (c) and (e)) and low spin (right panel: (b), (d) and (f)) ground state. (a) and (b) correspond to a polarization of the incident light parallel to the CuO$_2$-plane, whereas (c) and (d) show the XAS for incident light polarization perpendicular to the CuO$_2$-plane. In (e) and (f) the linear dichroism is shown. Blue thick lines represent the experimental data (equal in left and right panel) and red thin lines correspond to the result from multiplet ligand-field theory.}
	\label{fig:xasfig}
\end{figure}

XAS spectra as a function of the incident light polarization at the Ni L$_{2/3}$-edge for a sample with 1\% Ni doping are shown in \figref{fig:xasfig}. In panel (a) and (b) the polarization of the light is parallel to the CuO$_2$-planes (blue thick lines, parallel to the crystallographic $b$- or $c$-axis). This spectra is dominated by two peaks around 856 eV and 872 eV. In contrast, when the incident light polarization is aligned perpendicular to the CuO$_2$-planes (parallel to the $a$-axis), the XAS signal vanishes almost completely, as shown in \figref{fig:xasfig} (c) and (d). \figref{fig:xasfig} (e) and (f) show the so-called linear dichroism, which is given by the difference of the spectra measured at the two different polarizations and which is, hence, a measure of the anisotropy of the XAS signal. 
We now compare these experimental spectra to simulated XAS spectra based on multiplet ligand-field theory for a single square-planar Ni0$_4$ cluster. These calculations were done using the \textit{Quanty}~\footnote{\url{http://quanty.org/}} software package~\cite{Haverkort2012}. In the left panel of \figref{fig:xasfig} ((a), (c) and (e) red thin lines) we show the results of a calculation where the crystal field parameters and the hybridization between Ni and O was chosen such that the Ni realizes a high spin configuration, i.e., $S = 1$. In a second calculation, shown in the right panel of \figref{fig:xasfig} ((b), (d) and (f)), these parameters were adapted in order to yield a low-spin configuration, i.e., $S= 0$. \footnote{Within our multiplet ligand-field calculations the $D_{4h}$-symmetry was used to describe the square-planar CuO$_4$ cluster. Accordingly, the crystal-field is described by three parameters: $10D_q, D_t$ and $D_s$ and the ligand-field is modeled by one parameter $pd_{\sigma x}$ which corresponds to the hopping amplitude between Ni and O within the plane. The hopping in the perpendicular direction $pd_{\sigma z}$ is zero. For the two  calculations we used the following parameters: For the low spin state the crystal-field parameters were set to $10D_q = 0.76$\,eV, $D_t= 0$\,eV and $D_s = 0.4$\,eV and the ligand-field parameter were set to $pd_{\sigma x} = -2$\,eV. For the high spin state the parameters were: $10D_q = 0.76$\,eV, $D_t = 0$\,eV,  $D_s = 0.16$, and $pd_{\sigma x} = -1.4$\,eV.} 
Apparently, the high-spin calculation fails to reproduce the strong anisotropy of the XAS and, in particular, the almost vanishing signal for out-of-plane polarization (c.f. \figref{fig:xasfig} (c)). Contrary, the low-spin calculation correctly predicts the experimentally observed XAS spectra (c.f. \figref{fig:xasfig} (b) and (d)). Particularly, in \figref{fig:xasfig} (f) it can be seen that
theory and experiment agree even on a quantitative level. We also found that the qualitative shape of the simulated XAS spectra does not depend on details of the parameter set: For different parameter sets realizing a low spin (high spin) state, the simulated spectra look qualitatively similar to the spectra shown in the right (left) panels of \figref{fig:xasfig}. 
Accordingly, our XAS data verify that the Ni impurities in $\mathrm{Sr Cu_{1-x} Ni_x O_2}$ are indeed in a low spin state.  

The low-spin ($S\!=\!0$) ground state inferred from the combined XAS and NMR data is additionally supported by {\it ab initio} quantum chemistry calculations for divalent Ni-ion impurities within the SrCuO$_2$ lattice. These were performed on a fragment of one NiO$_4$ plaquette and the adjacent Cu and Sr sites with Madelung-field embedding, using the complete-active-space self-consistent-field (CASSCF) method and a subsequent post-CASSCF correlation treatment based on second-order perturbation theory \cite{QC_book_00}. The results obtained in the final step of the quantum chemistry study, also referred to as CASPT2 \cite{QC_book_00}, indicate a singlet $t_{2g}^6d_{z^2}^2$ ground state, with the  $t_{2g}^6d_{z^2}^1d_{x^2-y^2}^1$ triplet lying at 70 meV higher relative energy. 
Interestingly, the zeroth-order CASSCF treatment favors the triplet as ground state of the Ni$^{2+}$ ion, which is reminiscent of the competition and fine balance involving the different low-lying spin states of the Co$^{3+}$ ions in LaCoO$_3$ \cite{LaCoO_hozoi_08}. In other words, a $S\!=\!0$ ground state is only found after accounting for electron correlation effects beyond the CASSCF level; the largest contribution in stabilizing the lowest-spin state arises from ligand-to-metal charge-transfer processes \cite{LaCoO_hozoi_08,dd_CuO2_hozoi_11}. The results discussed here were obtained using lattice parameters as reported in Ref.\,\onlinecite{HeinauSrCuO2} and the quantum chemistry package {\sc molcas}\footnote{\textsc{molcas 7}, Department of Theoretical Chemistry, University of Lund, Sweden.}, with atomic-natural-orbital basis sets \cite{QC_book_00} from the standard {\sc molcas} library --- of quadruple-zeta quality for Ni and triple-zeta quality for O. The Cu$^{2+}$ and Sr$^{2+}$ nearest neighbors were modeled by total-ion potentials \cite{TIPs_CuSr} while the farther crystalline surroundings entered the calculations as an effective Madelung field corresponding to a fully ionic picture. 

\subsubsection{Spin-Lattice Relaxation}
\label{sec:NiT1}
\begin{figure}      
	\centering
	\includegraphics[width=\columnwidth]{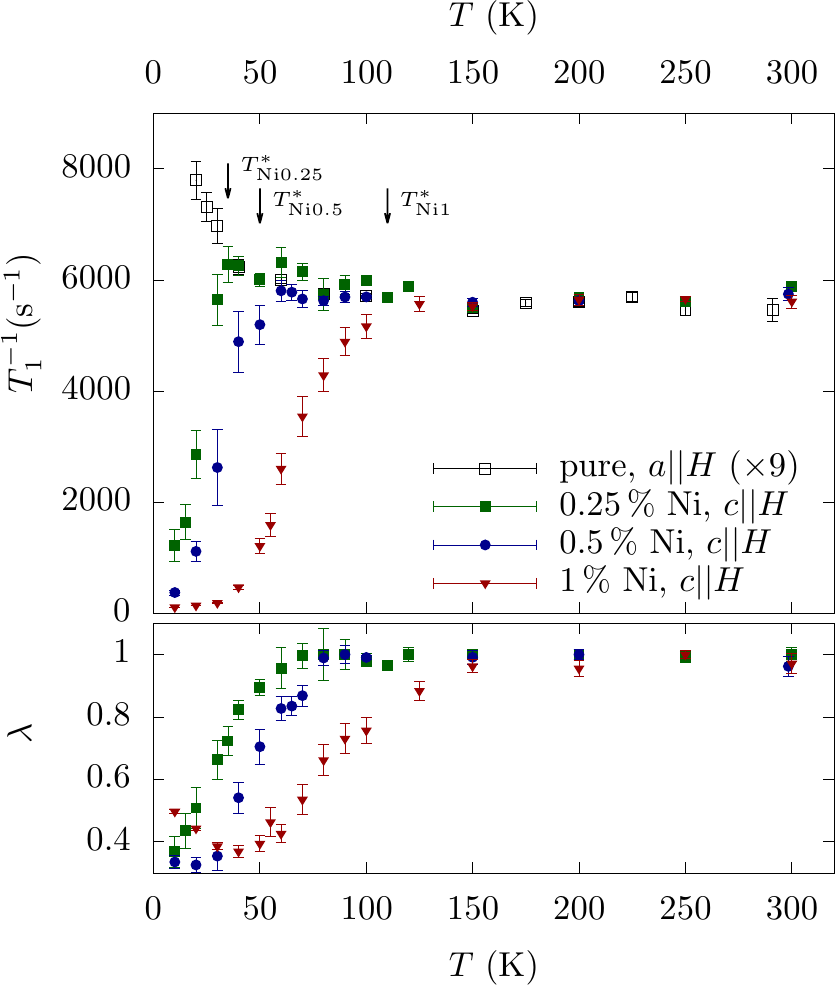}
	\caption{Spin-lattice relaxation rates $T_1^{-1}$ and stretching exponents $\lambda$ of \sNi for varying $x$, and spin-lattice relaxation rates $T_1^{-1}$ of pure SrCuO$_2$ 
		for comparison, each measured on the center of the $^{63}$Cu high-field satellite. While all doped samples were measured with the magnetic field $\mu_0 H\approx \SI{7}{T}$ parallel to the crystallographic $c$ axis (parallel to the chains), for $x=0.01$ data for the other two directions is also shown and the values of the pure compound were obtained with the magnetic field parallel to the $a$ axis (perpendicular to the chains). $T_1^{-1}$ values measured with $a||H$ are scaled. The black arrows indicate the crossover temperatures $T^\ast$.}
	\label{fig:relrat_SrCuO2+Ni}
\end{figure}

\figref{fig:relrat_SrCuO2+Ni} shows the results of the spin-lattice relaxation measurements on the Ni-doped samples compared to the pure compound. Spin-lattice relaxation rates and stretching exponents $\lambda$ are plotted over temperature. The measurements have been performed with the magnetic field $H$ parallel to the crystallographic $c$ axis for the Ni-doped samples, while the pure sample was measured with the magnetic field $H$ parallel to the crystallographic $a$ axis. Therefore, its absolute $T_1^{-1}$ values differ due to the differing hyperfine couplings. To allow for comparison, the relaxation rates of the pure compound are, thus, scaled by a factor of 9.

At high temperatures, $T_1^{-1}$ follows for all doping levels the behavior of the pure compound which is constant over a wide temperature range as it is theoretically expected for antiferromagnetic \mbox{$S=1/2$} Heisenberg chains \cite{Sachdev1994, Sandvik1995} and as it has already been experimentally verified earlier on SrCuO$_2$ and on the closely related spin chain compound Sr$_2$CuO$_3$ \cite{Hammerath2011a,Takigawa1996}.
Below a certain crossover temperature, which is $T^\ast_{\mathrm{N}0.25}\approx\SI{35}{K}$, $T^\ast_{\mathrm{N}0.5}\approx\SI{50}{K}$, and $T^\ast_{\mathrm{N}1}\approx\SI{110}{K}$ for N0.25, N0.5, and, N1, respectively,  $T_1^{-1}$ shows a strong decrease toward low temperatures.
The decrease of $T_1^{-1}$ is accompanied by a decrease of the stretching exponent $\lambda$ and thus by a growing spatial distribution of spin-lattice relaxation rates, which
levels off at lower temperatures. 

Due to the hyperfine coupling $A_\perp$ between nuclei and electrons, $T_1^{-1}$ measures the imaginary part of the dynamic spin susceptibility $\chi^{\prime\prime}$ of the electronic spin system at the NMR frequency.
For pure magnetic relaxation, it is given by 
\begin{equation}
T_1^{-1} \propto T \sum_{\vec{q}}A^2_\perp (\vec{q},\omega ) \frac{\chi^{\prime\prime}(\vec{q},\omega)}{\omega} \, .
\end{equation}
On a more intuitive level, the relaxation mechanism can be described as the scattering of thermally excited spinons by the copper nuclei \cite{Magishi1998}.

Thus, the decrease in spin-lattice relaxation rates clearly indicates the depletion of low-lying states in the spin excitation spectrum, and therefore points toward a spin gap. However, the distribution of spin-lattice relaxation rates, as indicated by $\lambda < 1$, implies that this spin gap varies spatially and can be characterized as a spin pseudogap on a macroscopic scale.

Usually, the magnitude of a spin gap is estimated by fitting the temperature dependence of $T_1^{-1}$ to an activated behavior \cite{Hammerath2011a,Takigawa1998,Ishida1994,Ohama1997, Imai1998} and using the activation energy as an estimate for the spin gap.
However, in our case, the spin-lattice relaxation rates do not decrease exponentially. This can be attributed to the spatial distribution of spin gaps, because the fast relaxation stemming from nuclei exposed to small gaps will dominate the recovery process at low temperatures. We use the crossover temperature $T^\ast$ as an estimate for the average gap energy, instead. The value of $T^\ast_{\mathrm{N}1} \approx \SI{110}{K}$ is close to the reported spin pseudogap $\Delta\approx\SI{90}{K}$  for $\SI{1}{\percent}$ of nickel doping,\cite{Simutis2013} which verifies $T^\ast$ to be a good estimate for the average gap magnitude. $T^\ast$ is almost proportional to the doping level. Therefore, we conclude that the average gap is proportional to the doping level, too. 
This is in agreement with the assumption that the individual chain segments show gaps $\Delta \propto 1/l$  and thus evidences the finite-size character of the spin pseudogap \cite{Eggert1992}.
The value of $T^\ast_{\mathrm{N}1} \approx \SI{110}{K}$ agrees to the crossover temperature of the single chain compound Sr$_2$CuO$_3$ doped with \SI{1}{\percent} of Ni doping \cite{Utz2015}, which shows that the double chain structure is not relevant for the gapping mechanism, similar to what has been observed in the Ca-doped variants of SrCuO$_2$ and Sr$_2$CuO$_3$ \cite{Hammerath2011a, Hammerath2014}.

\subsubsection{Frequency-Dependent Spin-Lattice Relaxation}
\label{sec:Ni_freq_dep_T1}
\begin{figure}  
	\subfloat[\label{fig:freqdepT1_SrCuO2_Ni1}]{%
   		\parbox{0.49\columnwidth}{\includegraphics{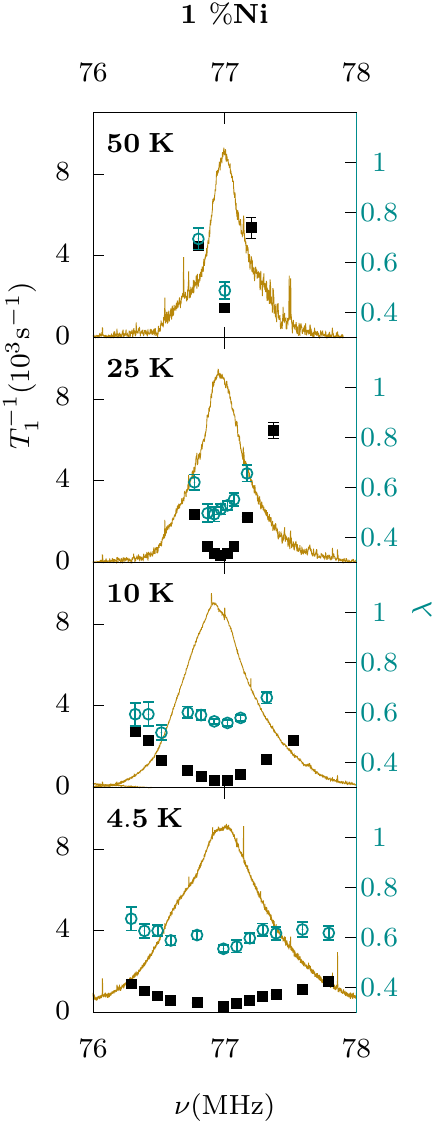}}
   	}
   	\subfloat[\label{fig:freqdepT1_SrCuO2_Zn2}]{%
   		\parbox{0.49\columnwidth}{\includegraphics{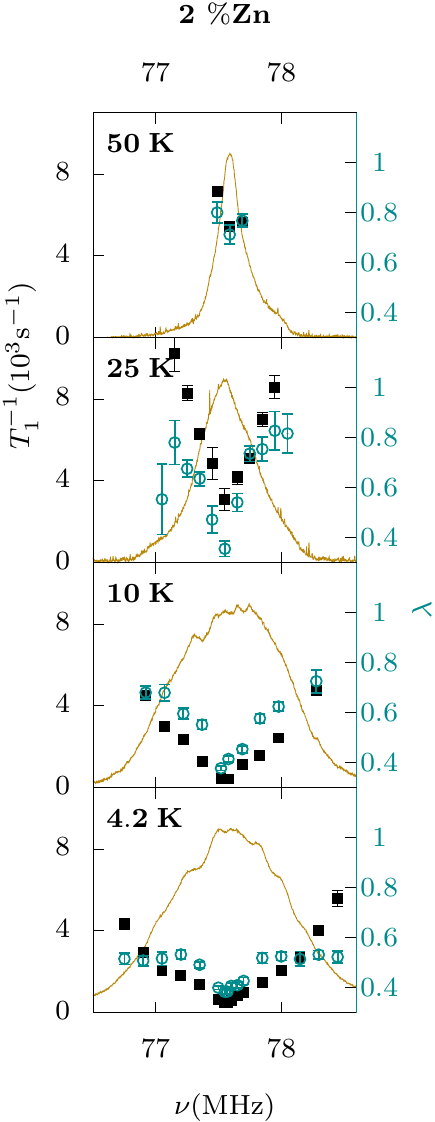}}
   	}
   	\caption{Frequency-dependent spin-lattice relaxation measurement on the $^{63}$Cu high-field satellite of SrCu$_{0.99}$Ni$_{0.01}$O$_2$ (a) and of SrCu$_{0.99}$Zn2$_{0.02}$O$_2$ (b) at different temperatures. Spin-lattice relaxation rates $T_1^{-1}$ (filled black squares), stretching exponents $\lambda$ (open cyan circles) and spectrum (dark gold line) were measured with the  $b$ axis parallel to the field $\mu_0 H = \SI{6.9981}{T}$ (a) and  $\mu_0 H = \SI{7.0488}{T}$ (b). The spectral intensity is normalized to the maximum of the high-field satellite and its scale is not shown in the graph.}
   	\label{fig:freqdepT1}
   \end{figure}

To gain further knowledge about the spatial variation of spin gaps, the frequency dependence of spin-lattice relaxation within the broad resonance lines has been studied.
\figref{fig:freqdepT1_SrCuO2_Ni1} shows spin-lattice relaxation rates $T_1^{-1}$ and stretching exponents $\lambda$ measured at different positions within the $^{63}$Cu high-field satellite of N1 at various temperatures.
Thereby, the upper limit of the temperature series was determined by practical considerations concerning the temperature-dependent width of the resonance lines, which should be large enough to allow for several spin-lattice relaxation measurements with reasonable spacing in between.

Within the studied temperature range a strong frequency dependence of $T_1^{-1}$ and $\lambda$ is observable. 
Spin-lattice relaxation rates $T_1^{-1}$ at all positions decrease towards low temperatures. However, this decrease is less steep the larger the distance to the center of the resonance line. Therefore, $T_1^{-1}$ at all temperatures is smallest in the center of the line and largest at its edge. However, with decreasing temperature, the differences get smaller again, as  $T_1^{-1}$ at the edge approaches zero, too.
$\lambda$ is minimal at the center and larger at the outer parts of the resonance lines. Its frequency dependence also flattens towards low temperatures.

Putting everything together one can conclude that Cu nuclei which contribute to the outer parts of the resonance lines probe a narrow distribution of small spin gaps, while Cu nuclei contributing to the center of the resonance lines probe a broad distribution of large and small spin gaps.
In principle, there are two possibilities on how the gap could vary to get such a frequency dependence in combination with the LAM. One possibility is that the gap varies within single chain segments in a way that it is small at sites where the local susceptibility is large and large at sites where the local susceptibility is close to zero. Assuming a one-to-one correspondence between local susceptibility and gap, this idea can easily be disproved: it would mean a well-defined $T_1^{-1}$ value at every point in the spectrum and $\lambda < 1$ would not occur.
The results from the preceding section suggest another interpretation. The doping dependence of the spin-lattice relaxation measurements showed that the magnitude of the spin gap depends on the chain length --- they are inversely proportional to each other.
Moreover, the analysis of the spectra showed that the shape of the LAM depends on the chain length, too. The amplitude of the local susceptibility at low temperatures of even segments is smaller the shorter its length. In fact, both dependencies are intimately related according to the model of finite chain segments.
These conclusions fit remarkably well to the frequency dependence of spin-lattice relaxation. Upon cooling down, the temperature reaches at first values comparable to the size of the gap of the shortest chain segments. Two things are happening then during cooling: the relaxation rate of these segments starts to decrease and their local susceptibility is suppressed, which means that their contribution to the spectral intensity is rearranged toward the center of the resonance line. By further cooling the sample, longer chain segments start to take part in this process. Chain segments, whose local susceptibility is not suppressed yet and whose spin-lattice relaxation corresponds still to the high-temperature behavior, contribute to the full width of the resonance line. The shorter the chain segment and, therefore, the smaller its corresponding spin-lattice relaxation rate, the narrower is the region around the center where this segment contributes to the resonance line. Thus, one finds a broad distribution of large and small relaxation rates close to the center of the resonance line, while far away from the center there is a narrow distribution of large relaxation rates only.

Note that an additional variation of $T_1^{-1}$ within one chain segment cannot be excluded. There is no reason to assume that the imaginary part of the dynamic susceptibility $\chi^{\prime\prime}$ in such a translational invariant system is homogeneous. The static susceptibility varies within one chain segment too, as one can see in the NMR spectra. Nevertheless, there is no possibility to prove the possible variation of $T_1^{-1}$ within one chain segment, as long as a complete model for the local static susceptibility, i.e. for the spectra, is missing.

\subsubsection{Summary}
In this section, Cu NMR spectra and spin-lattice relaxation measurements of Ni-doped SrCuO$_2$ have been presented and discussed.
The temperature and doping dependence of the spectra, the temperature dependence of spin-lattice relaxation and its variation within the broad resonance lines as well can basically be understood using the model of finite chain segments. Therefore, the results strongly indicate that nickel impurities behave as scalar defects in the cuprate spin chains. Thereby, the observed behavior is in all respects the same as for the Ni-doped single chain compound \cite{Utz2015}. For \SI{1}{\percent} of Ni doping even the size of the gap is the same as the crossover temperatures $T^*_{N1}$ coincide. 
This confirms that the double chain structure is not relevant and that the single chain model is suitable to describe the magnetic behavior of SrCuO$_2$ even in the case of in-chain doping.
NMR spectra have been simulated based on the model of finite chain segments and have been compared to the measured spectra. Even though the simulation catches the essential features of the measurements, there are deviations.
The measured low-temperature spectra are much broader than the simulations.
The reasons for this is most likely the interchain coupling, which is not considered in the model calculations. Finally, XAS spectra have been measured for a 1\% doped sample and have been compared to simulated XAS spectra based on multiplet ligand-field theory. This comparison as well as quantum chemistry calculations for divalent Ni-ion impurities within the SrCuO$_2$ lattice confirm the $S=0$ low spin state of the Ni impurities.  

\subsection{Palladium-Doped Samples}
\label{sec:PdDoping}
Palladium is known to be a $S=0$ impurity \cite{Sirker2007, Kojima2004, Kawamata2008}. NMR measurements on Pd-doped samples thus provide further evidence that the nickel impurities act as scalar defects. Therefore, spectra and spin-lattice relaxation on the center of the $^{63}$Cu mainline have been measured on P1 for various temperatures under the same conditions as for N1.

\begin{figure} 
	\centering
	\includegraphics{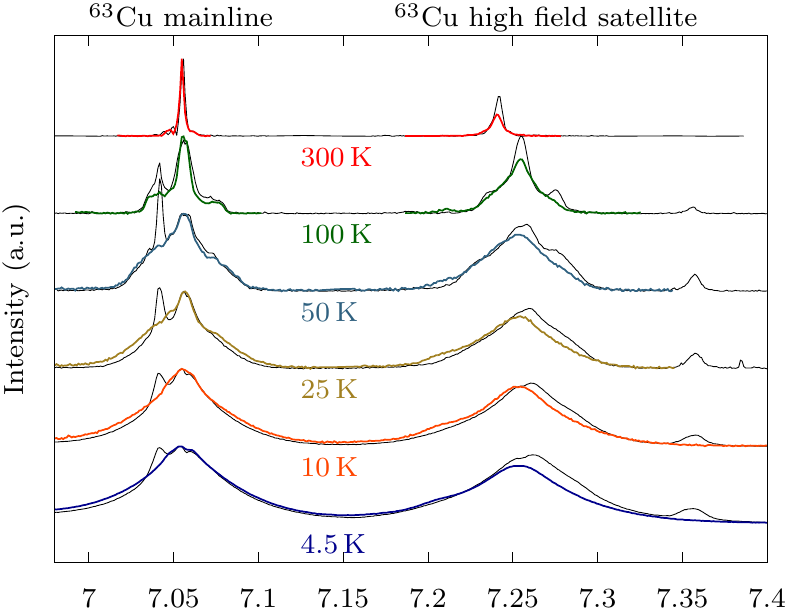}
	\caption{NMR spectra of SrCu$_{0.99}$Pd$_{0.01}$O$_2$ at different temperatures measured with a fixed frequency of $\SI{80}{MHz}$ by varying the magnetic field  $H||b$ (perpendicular to the chains). The spectra of SrCu$_{0.99}$Ni$_{0.01}$O$_2$ are plotted as black lines for comparison.}
	\label{fig:SrCuO2_PdNi}
\end{figure}

The spectra are plotted in \figref{fig:SrCuO2_PdNi} and compared to the ones of N1. One can see that they are almost exactly congruent. The main difference is the absence of the additional peaks as in the Ni-doped case. Thus, Pd doping does not lead to local lattice distortions. Further differences can be observed on the $^{63}$Cu high-field satellite.  On the one hand its center is slightly shifted. The reason for this is a small but almost unavoidable misalignment of the sample.\footnote{The position of the satellite lines shows a much stronger angular dependence than the position of the mainline because the satellite lines are affected by the quadrupolar shift.} On the other hand, the satellite line shows less structure than the mainline. Especially at \SI{100}{K} the shoulder features are barely visible on the satellite line of P1 while they are well resolved for N1. As this smearing only concerns the satellite lines, it must have quadrupolar origin and can also be explained by the small misalignment of the sample.

\begin{figure}      
	\centering
	\includegraphics{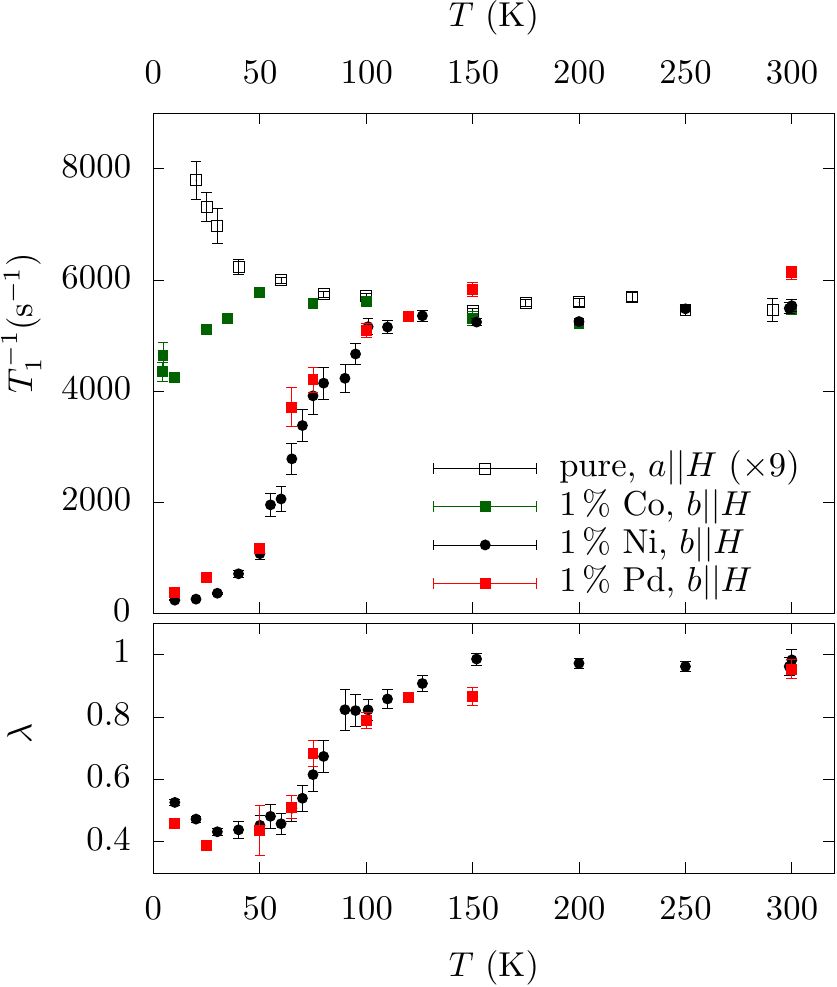}
	\caption{Spin-lattice relaxation rates $T_1^{-1}$ and stretching exponents $\lambda$ of SrCu$_{0.99}$Pd$_{0.01}$O$_2$ and SrCu$_{0.99}$Co$_{0.01}$O$_2$ and of SrCu$_{0.99}$Ni$_{0.01}$O$_2$ and pure SrCuO$_2$ for comparison, each measured on the center of the $^{63}$Cu high-field satellite. All samples were measured with the magnetic field $\mu_0 H\approx \SI{7}{T}$ parallel to the crystallographic $b$ axis (parallel to the chains), except of the pure sample which was measured  with the magnetic field parallel to the $a$ axis (perpendicular to the chains).}
	\label{fig:relrat_SrCuO2+PdCo}
\end{figure}

The spin-lattice relaxation measurement have been analyzed in the same way as the one on N1 (see \figref{fig:relrat_SrCuO2+Ni}). The resulting fitting parameters --- $T_1^{-1}$ and $\lambda$ --- are plotted in \figref{fig:relrat_SrCuO2+PdCo} together with the ones of N0, N1, and C1 for comparison. Both quantities follow very closely the behavior of N1.

As spectra as well as spin-lattice relaxation measurements reproduce essentially the corresponding observations on the Ni-doped sample, the reader is referred to \secref{sec:NiDoping} for a more thorough analysis and interpretation. The coincidence of the NMR results on both samples further proves that nickel acts as a spin 0 defect in SrCuO$_2$. The fact that even the spectra are essentially the same  suggests that nickel is in its low spin state, since a screened nickel $S=1$ should lead to a signature of the screening in the NMR spectra \cite{Rommer2000}. However, it might also be possible that the screening would have already saturated in the observed temperature range, depending on the impurity coupling $J_{imp}$, and is, therefore, not visible in the spectra \cite{Rommer2000}.

\subsection{Zinc-Doped Samples}
The Zn-doped samples were originally planed to serve as the necessary comparison for the Ni-doped case. However, they were not suitable for this purpose as the Zn ions turned out to avoid to occupy copper sites. NMR measurements supporting this conclusion are shown in this section.

\subsubsection{Spectra}
\begin{figure*}  
	\centering
	\includegraphics{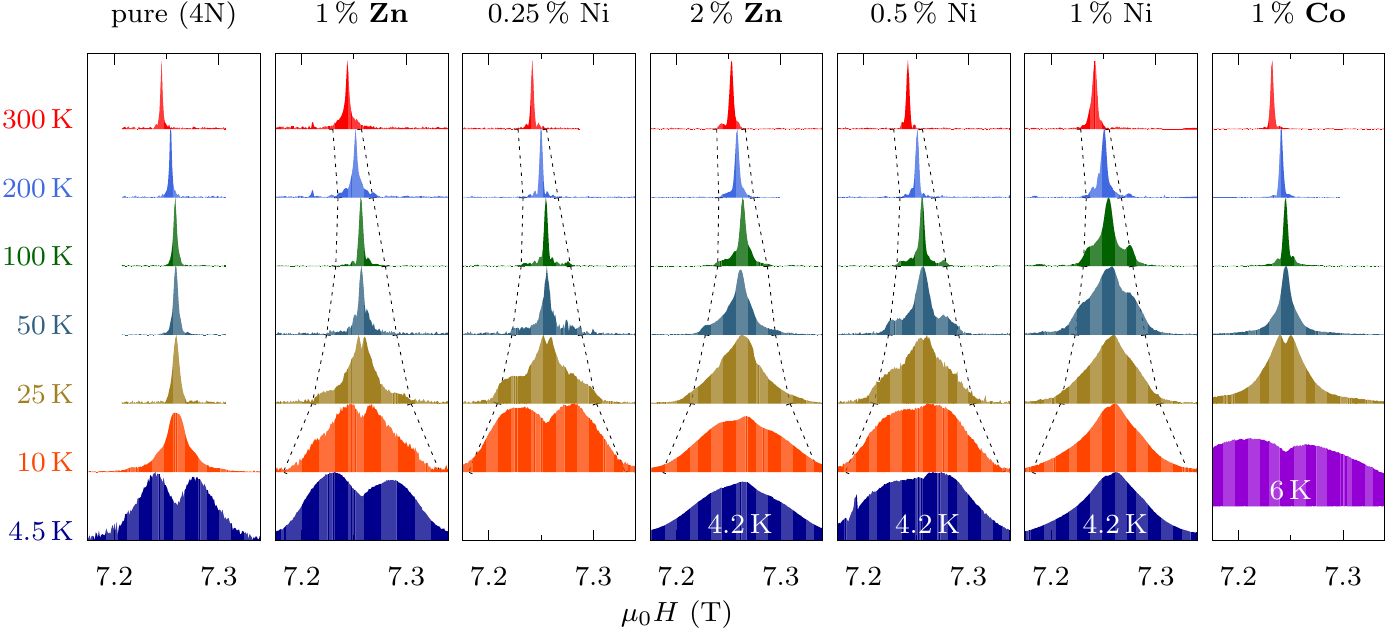}
	\caption{$^{63}$Cu high-field satellite of SrCuO$_2$, \sZn, \sNi and SrCu$_{0.99}$Co$_{0.01}$O$_2$ measured with a fixed frequency of $\SI{80}{MHz}$ by varying the magnetic field $H||b$ (perpendicular to the chains). The black dashed lines (more precise: the ends of their segments) indicate the $1/\sqrt(T)$-behavior, which is expected for the shoulder feature from the model of semi-infinite chains with \mbox{$\sqrt(T)\Delta H/2H_0 = 0.0326$}. The plots of the pure, Zn-doped and Ni-doped samples are arranged such that they show increasing impurity effects from the left to the right (see text).}
	\label{fig:highfieldsat_comp_SrCuO2_ZnCo}
\end{figure*}

\figref{fig:highfieldsat_comp_SrCuO2_ZnCo} shows the high-field satellites of Zn-doped SrCuO$_2$ together with the pure, Ni-doped and Co-doped versions for comparison.
Towards low temperatures, they broaden and obtain some structure. This process is again of magnetic origin, as mainline and satellites show the same behavior. An additional set of peaks as in the case of Ni doping does not show up.
The black dashed lines show the $1/\sqrt{T}$-behavior of the edges of the shoulder feature which is expected from the model of semi-infinite chains.
Also the Zn doped samples show shoulder features which follow essentially this behavior. However, the intensity of the shoulders is much smaller than the one of the corresponding Ni-doped samples. The model of semi-infinite chains predicts the intensity of the shoulder feature to be proportional to the chain break concentration. Therefore, the samples can be put in an order with increasing real defect concentration. This has been done in \figref{fig:highfieldsat_comp_SrCuO2_ZnCo} by the arrangement of the panel which reflects an increasing in-chain doping from the left to the right for the Zn- and Ni-doped samples. This order is particularly comprehensive by considering the spectra at \SI{50}{K} and \SI{100}{K}. One can see that the shoulder feature of nominal \SI{1}{\percent} Zn doping is even lower in intensity than of \SI{0.25}{\percent} Ni doping. The shoulder feature of nominal \SI{2}{\percent} Zn doping seems to be almost the same as for \SI{0.5}{\percent} Ni doping. This sequence is also reflected in the low-temperature spectra. However, the behavior is more complex. For example at \SI{10}{K} the spectra broaden at first with increasing defect concentration. They develop two broad humps. Then the spectra loose this humped structure and get narrower again. This behavior fits remarkably well into the picture drawn in \secref{sec:NiDoping}. Low doping levels lead to a proliferation of the local alternating susceptibility at low temperatures. Eventually it is even reflected in neighboring chains due to the interchain coupling. Anyhow, most of the Cu spins are polarized, which leads to the broad and humped lines. Higher doping levels result in shorter chain segments. As short even chains lock in the singlet state and short odd chains are not observable anymore, the lines get narrower with further increase of the doping level.
Thus, from the NMR spectra one can conclude in a consistent way that the in-chain defect concentration of Zn-doped SrCuO$_2$ is much lower than the nominal doping level. 
It seems to be smaller than \SI{0.25}{\percent} for \SI{1}{\percent} nominal Zn doping and almost \SI{0.5}{\percent} for \SI{2}{\percent} nominal Zn doping.

\subsubsection{Spin-Lattice Relaxation}
\begin{figure}      
	\centering
	\includegraphics{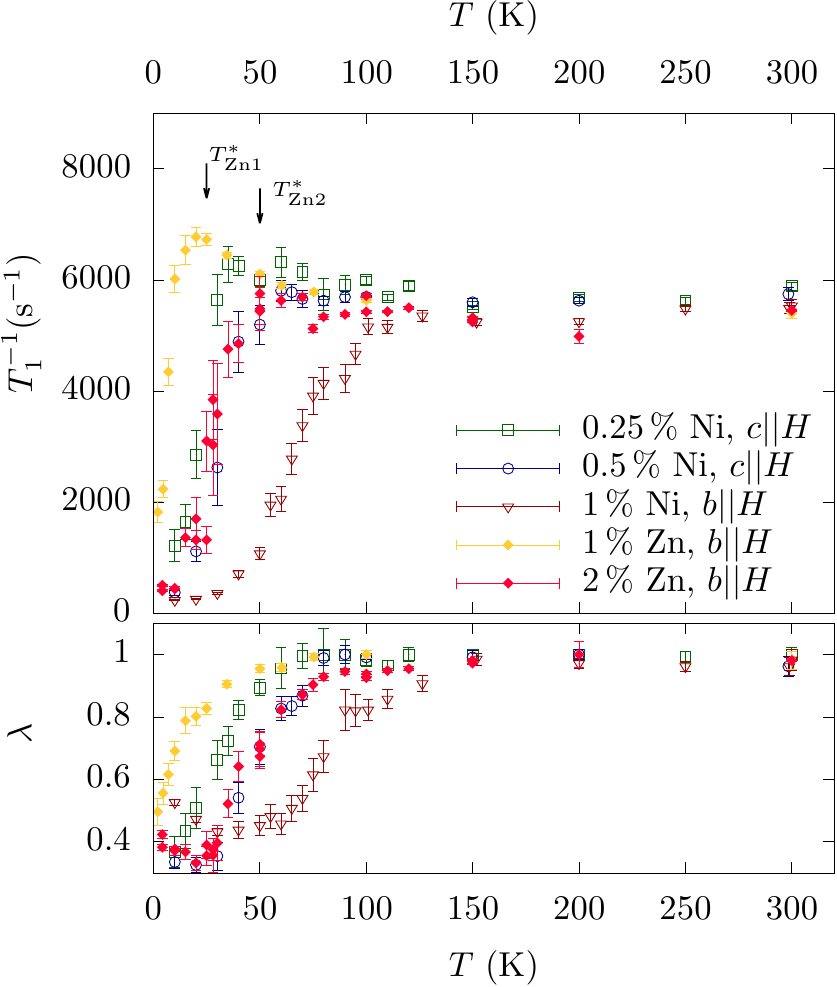}
	\caption{Spin-lattice relaxation rates $T_1^{-1}$ and stretching exponents $\lambda$ of \sZn for varying $x$ and of \sNi for comparison, each measured on the center of the $^{63}$Cu high-field satellite. All samples were measured with the magnetic field $\mu_0 H\approx \SI{7}{T}$ parallel to the crystallographic $b$ axis (parallel to the chains).
		The black arrows indicate the crossover temperatures $T^\ast$.}
	\label{fig:relrat_SrCuO2+Zn}
\end{figure}

\figref{fig:relrat_SrCuO2+Zn} shows the results of the spin-lattice relaxation measurements on  the Zn-doped samples compared to the pure and Ni-doped ones.
One can see that $T_1^{-1}$ of the  Zn-doped samples follows also the behavior of the pure compound down to a certain temperature $T^\ast$ and decreases strongly towards lower temperatures. This decrease of $T_1^{-1}$ is accompanied by a decrease of the stretching exponent $\lambda$. 
Thus, the behavior is essentially the same as for the Ni-doped samples.
It suggests that Zn doping also induces a distribution of spin gaps by cutting the chains into segments with finite length.
However, also this effect is much smaller than with nickel or palladium doping.
A nominal Zn content of $\SI{1}{\percent}$ leads to $T^\ast_{Z1}=\SI{25}{K}$ only, which is even smaller than $T^\ast_{N0.25}=\SI{35}{K}$ for \SI{0.25}{\percent} Ni doping.
A nominal Zn content of $\SI{2}{\percent}$ results in $T^\ast_{Z2}=\SI{50}{K}$, which coincides with $T^\ast_{N0.5}=\SI{50}{K}$ for \SI{0.5}{\percent} Ni doping.
According to the results in \secref{sec:NiT1}, $T^\ast$ should be proportional to the in-chain defect concentration.
Therefore, also the $T_1$ measurements indicate that the real defect concentration $c$ is much smaller than the nominal doping level. Moreover, the same sequence of increasing defect concentrations can be deduced: $c(Z1) < c(N0.25) < c(Z2) \approx c(N0.5) < c(N1)$.

\subsubsection{Frequency-Dependent Spin-Lattice Relaxation}
\figref{fig:freqdepT1_SrCuO2_Zn2} shows spin-lattice relaxation rates $T_1^{-1}$ and stretching exponents $\lambda$ measured at different positions within the $^{63}$Cu high-field satellite of Z2 for different temperatures between \SI{50}{K} and \SI{4.2}{K}.
As in the case of Ni doping, a strong frequency dependence of $T_1^{-1}$ and $\lambda$ can be observed.  $T_1^{-1}$ and $\lambda$ are smallest at the center of the resonance lines and get larger the further away from the center they were measured.
In \figref{fig:freqdepT1_SrCuO2_Zn2} one can see that at \SI{50}{K} --- the temperature where the spin-lattice relaxation rate at the center of the resonance line starts to drop --- the frequency dependence is not much pronounced. It intensifies toward low temperatures as it has been observed for the case of Ni doping (see \secref{sec:Ni_freq_dep_T1}).
However, there is one difference to the case of Ni doping. At the outer parts of the resonance line, relaxation rates of $T_1^{-1}>\SI{6000}{s^{-1}}$ are obtained. This is more than the high temperature value of $T_1^{-1}\approx \SI{5500}{s^{-1}}$.
The model of finite chains does not predict an increase of spin-lattice relaxation rates towards low temperatures. Thus, there has to be an additional effect.
An upturn toward low temperatures is also observable for the spin-lattice relaxation rates measured on the center of the resonance line of pure SrCuO$_2$ (see \figref{fig:relrat_SrCuO2+Ni}), which is most probably  a manifestation of critical fluctuations associated with the nearby phase transition to 3D ordering at $T_N \approx \SI{2}{K}$ \cite{Zaliznyak1999, Matsuda1995}, i.e. an additional effect due to the interchain coupling.
So, it might be that the critical fluctuations are still effective for long chain segments which contribute to the outer parts of the resonance lines,  whose spin-lattice relaxation rate is, thus, not suppressed yet.
This would also explain why this increase in spin-lattice relaxation rate is not observed for \SI{1}{\percent} of Ni doping. There, the suppression of $T_1^{-1}$ due to the finite chain lengths concerns already the full width of the resonance line before the temperature range with critical fluctuations is reached. The reasons are: on the one hand the decrease of $T_1^{-1}$ due to the gap occurs at higher temperatures and on the other hand $T_N$ is suppressed by the doping \cite{Karmakar2015b}.

\subsubsection{Chemical Analysis}
To determine the actual zinc content, the samples have been studied by the ICP-OES (Inductively Coupled Plasma Optical Emission Spectroscopy) method. For this method the samples are ground and dissolved by a suitable solvent. Then the material is brought into an argon plasma and the optical emission of the elements is analyzed. For comparison, the Ni-doped samples have also been examined.
\tabref{tab:ICP} shows the results. One can see that the actual zinc content is in fact much smaller than the nominal doping level, whereas for nickel, actual and nominal doping level agree well.
However, the actual Zn content is still much higher than expected from the NMR measurements and their comparison with the Ni-doped samples. As shown in the last sections, one would expect the sample Z1 to contain less than \SI{0.25}{\percent} of Zn and the sample Z2 to contain in about \SI{0.5}{\percent} of Zn.
The reason for this discrepancy is that the chemical analysis measures the total zinc content, while the analysis of the NMR measurements is only susceptible to chain breaks.
Therefore, one can conclude that either only a fraction of the contained Zn occupies copper sites or that the Zn impurities cluster in the chain, such that several Zn ions are responsible for one chain break.
Thus, the zinc used in the growth process does not only avoid being incorporated into the sample but also avoids occupying the copper site or occupies with high probability consecutive sites.

\begin{table}
	\centering
	\resizebox{0.8\columnwidth}{!}{%
		\begin{tabular}{|l|l|l|}
			\hline  \textbf{Sample} & \textbf{nominal doping level} & \textbf{measured doping level}  \\ 
			\hline  Zn1	&  \SI{1}{\percent} &  \SI{0.357 \pm 0.001}{\percent}\\ 
			\hline  Zn2 &  \SI{2}{\percent}&  \SI{1.493 \pm 0.008}{\percent}\\  
			\hline  Ni0.25 &  \SI{0.25}{\percent}&  \SI{0.268 \pm 0.002}{\percent}\\ 
			\hline  Ni0.5 &  \SI{0.5}{\percent}&  \SI{0.437 \pm 0.002}{\percent}\\ 
			\hline  Ni1 &  \SI{1}{\percent}&  \SI{0.996 \pm 0.004}{\percent}\\ 
			\hline
		\end{tabular}} 	
		\vspace{5mm}
		\caption{Real doping levels of the Ni- and Zn-doped samples obtained by the ICP-OES method.}
		\label{tab:ICP}
	\end{table}

\subsubsection{Summary}
In this section, NMR measurements and a chemical analysis of the Zn-doped samples are presented.
All NMR measurements agree well with the model of finite chain segments, which confirms that Zn indeed produces chain breaks which lead to finite size gaps and a LAM as in the case of Ni- and Pd-doping.
However, these measurements show in comparison with the Ni-doped samples that the actual in-chain impurity content is much smaller than the nominal one. The ICP-OES measurements reveal an overall Zn content which is indeed smaller than the nominal one, but still much larger than the observed in-chain impurity content. This means that not all of the contained Zn ions replace copper ions or that Zn clusters in the chain.
This agrees to recent susceptibility measurements on Zn-doped Sr$_2$CuO$_3$ which suggest that not all of the zinc occupies copper sites \cite{Karmakar2015b} and to the observation that Zn accumulates in the floating zone during the growth of single crystals of Zn-doped Sr$_2$CuO$_3$ \cite{Karmakar2015c}. 

\subsection{Cobalt-Doped Samples}
In contrast to all other dopants discussed in this work, Co is most certainly a magnetic impurity. The spin state is not clear, but it is surely half-integer. Due to the square planar arrangement of the oxygen ions, the Co$^{2+}$ ion could either be in a low spin state with $S=\frac{1}{2}$ or a high spin state with $S=\frac{3}{2}$. Therefore, the behavior is expected to differ from the model of finite chain segments.

\figref{fig:highfieldsat_comp_SrCuO2_ZnCo} shows the $^{63}$Cu high-field satellite of SrCuO$_2$ doped with $\SI{1}{\percent}$ of Co together with the pure, Ni-doped and Zn-doped versions for comparison.
In contrast to the other dopings, there is hardly any broadening of the resonance line down to $T=\SI{100}{K}$. The broad shoulder feature corresponding to the maxima of the LAM at open chain ends, does not show up. From \SI{50}{K}  down, the resonance line starts to broaden and at \SI{25}{K} a splitting of the tip is visible. Both evidence growing antiferromagnetic correlations in the system. At \SI{6}{K}, there is intensity almost everywhere in the spectrum. The dip in intensity at the center shows that the number of sites with zero local magnetization are degraded. This agrees well with the observation of 3D magnetic order just below \SI{6}{K} \cite{Karmakar2017}. At lower temperatures, it was not possible to obtain spectra due to the very fast spin-spin relaxation rates. This can be explained as a result of the strongly varying local magnetic field in the short-range ordered phase \cite{Karmakar2017}.

The spin-lattice relaxation rates $T_1^{-1}$ measured at the center of the $^{63}$Cu high-field satellite of C1 are shown in \figref{fig:relrat_SrCuO2+PdCo} together with the ones of N0, N1, and P1 for comparison. The stretching parameter $\lambda$ is not plotted for C1 as the recovery curves of nuclear magnetization could be fit with the usual relaxation function, which corresponds to $\lambda = 1$ for the entire temperature range.
One can see that $T_1^{-1}$ follows closely the behavior of the pure compound down to \SI{50}{K}, which is the same temperature where a broadening of the spectrum is first observed. Towards lower temperatures, $T_1^{-1}$ decreases, but only slightly. The decrease differs considerably from the behavior of Ni-doped, Pd-doped, and Zn-doped samples and $T_1^{-1}$ does not go to zero. Thus, a spin gap seems not to be present in the case of Co-doping.

Hence, neither the typical LAM nor the opening of a spin gap are observed upon Co-doping. This means there is no sign of chain segmentation. Instead, the antiferromagnetic correlations are increased. This agrees well to recent results on inelastic neutron scattering and $\mu$SR measurements\cite{Karmakar2017}. They show the absence of a gap and the increase of the ordering temperature.
It is still an open question how Co doping increases the tendency to order, but it is presumably related to the single ion anisotropy of the Co$^{2+}$ ions \cite{Karmakar2017, Bera2014, Sati2006, Slonczewski1961}.

\section{Conclusions}
We studied the cuprate spin chain system SrCuO$_2$ intentionally doped with Ni, Pd, Zn, and Co impurities by means of nuclear magnetic resonance (NMR).
For all samples, Cu NMR spectra have been obtained and Cu NMR spin-lattice relaxation measurements have been performed within a wide temperature range from \SI{300}{K} to \SI{4.2}{K}.\\

\textbf{Scalar Impurities}\\
The NMR spectra of the Ni-, Pd-, and Zn-doped samples show a characteristic broadening with decreasing temperature, which reacts very sensitively on slight doping differences.
The spin-lattice relaxation measurements of the same samples reveal a broadening distribution of decreasing spin-lattice relaxation rates upon cooling down, which vary within the broad resonance lines.
These measurements can be essentially understood using the model of finite segments of the spin $1/2$ antiferromagnetic Heisenberg chain. This means that Ni, Pd, and Zn impurities seem to simply cut the infinitely long chains into segments with random lengths.

Another major result concerns the role of nickel impurities.
While for the 2D cuprates nickel impurities are known to be in the high-spin state ($S=1$), this was not clear for the chain cuprates.
The comparison of the NMR spectra and the spin-lattice relaxation measurements of a sample of Ni-doped SrCuO$_2$ with the corresponding measurements of a Pd-doped sample with the same impurity concentration shows that  both impurities lead to the same behavior. The strong agreement of the NMR results on Pd- and Ni-doped samples strongly suggests that Ni impurities in SrCuO$_2$ adopt the low spin state ($S=0$) and, therefore, act as native spin 0 impurities in contrast to their behavior in the 2D cuprates. Moreover, the low spin state of the Ni impurity has been confirmed by XAS measurements and the comparison to calculated spectra, and by quantum chemistry calculations. 
Also the case of Zn-doping shows peculiarities. By comparing the measurements on the Zn-doped samples with the Ni-doped case, it could be shown that either only a fraction of the contained Zn occupies copper sites or Zn clusters in the chain. Zinc is therefore not a good dopant for the intentional evocation of chain breaks in SrCuO$_2$.
Moreover, the experiments showed that even in the case of in-chain doping there are no essential differences between the behavior of SrCuO$_2$ and Sr$_2$CuO$_3$ \cite{Utz2015}. This confirmed once again that the single chain model is suitable to describe the magnetic behavior of SrCuO$_2$ in spite of its double chain structure.

Further improvement and confirmation of the obtained insights could become possible with a complete model of the low-temperature spectra taking into account the interchain couplings. This would allow to simulate the variation of spin-lattice relaxation measurements within the broad resonance lines and, therefore, also the distribution of spin-lattice relaxation rates at the center.\\

\textbf{Magnetic Impurities}\\
Doping with the magnetic impurity Co shows a substantially different behavior than the other cases. There are no signs of finite size gaps or of a LAM, which shows that chain segmentation does not occur due to the half-integer spin state of the Co$^{2+}$ ions. Instead, there is evidence for enhanced antiferromagnetic correlations, which agrees with the observation of the increased ordering temperature \cite{Karmakar2017}.

\begin{acknowledgments}
The authors thank S. Nishimoto and S.-L. Drechsler for discussion and Andrea Vo{\ss} for conducting the ICP-OES measurements.
This work has been supported by the European Commission through the LOTHERM project (Project No. PITN-GA-2009-238475) and by the Deu\-tsche For\-schungs\-ge\-mein\-schaft (DFG) through Grant No. GR3330/4-1, through the D-A-CH project No. HE3439/12 and through the Son\-der\-for\-schungs\-be\-reich (SFB) No. 1143. 
\end{acknowledgments}

\end{document}